\documentclass[aps,showpacs,twocolumn,pra,superscriptaddress]{revtex4-1}
\usepackage[english]{babel}
\usepackage[latin1]{inputenc}
\usepackage{hyperref}
\usepackage{amsmath, amssymb, amsfonts}
\usepackage{amssymb,xcolor}
\usepackage{graphicx}
\usepackage{multirow}
\graphicspath{{images/}}
\usepackage{exscale}
\usepackage{graphicx}
\usepackage{latexsym}
\newcommand{\ket}[1]{|#1\rangle}
\newcommand{\bra}[1]{\langle #1|}

\newcommand{\Cr}{\mathcal{C}_{r}}

\newcommand{\D}{\mathcal{D}}
\newcommand{\I}{\mathcal{I}}
\newcommand{\C}{\mathcal{C}}
\newcommand{\CC}{\mathcal{CC}}

\newcommand{\h}{\mathcal{H}}
\newcommand{\n}{\nonumber\\}
\newtheorem{theorem}{Theorem}

\begin{document}

\title{The role of coherence during classical and quantum decoherence}
\author{Jin-Xing Hou}
\affiliation{Institute of Modern Physics, Northwest University, Xi'an 710069, China}
\affiliation{Shaanxi Key Laboratory for Theoretical Physics Frontiers, Xi'an 710069, China}
\affiliation{School of Physics, Northwest University, Xi'an 710069, China}
\author{Si-Yuan Liu}\email{syliu@iphy.ac.cn}
\affiliation{Institute of Modern Physics, Northwest University, Xi'an 710069, China}
\affiliation{Shaanxi Key Laboratory for Theoretical Physics Frontiers, Xi'an 710069, China}
\author{Xiao-Hui Wang}
\affiliation{School of Physics, Northwest University, Xi'an 710069, China}
\affiliation{Shaanxi Key Laboratory for Theoretical Physics Frontiers, Xi'an 710069, China}
\author{Wen-Li Yang}
\affiliation{Institute of Modern Physics, Northwest University, Xi'an 710069, China}
\affiliation{Shaanxi Key Laboratory for Theoretical Physics Frontiers, Xi'an 710069, China}

\begin{abstract}

The total correlation in a bipartite
quantum system is measured by the quantum
mutual information $\mathcal{I}$, which consists of quantum discord and classical correlation.
However, recent results in quantum information show that coherence, which is a part of total correlation, is more general and more fundamental.
The role of coherence in quantum resource theory is worthwhile to investigate.
We first study the relationship between quantum discord and coherence by decreasing the difference between them.
Then, we consider the dynamics of quantum discord, classical correlation and quantum coherence under incoherent quantum channels.
It is found that coherence indicates the behavior of quantum discord (classical correlation) for times $t<\bar t$, and indicates the behavior of classical correlation (quantum discord) for times $t>\bar t$. Moreover, the coherence frozen and decay characterize the quantum discord and classical correlation frozen and decay respectively.

\end{abstract}

\maketitle
\section{introduction}

As key resources of quantum information processing, quantum correlations can be used to achieve different tasks with different forms, such as
entanglement, quantum discord, coherence and so on.
It is widely accepted that quantum mutual information, which consists of classical and quantum correlations, is the measure of total correlation in bipartite quantum system.
Aiming to capture the total nonclassical correlation,
Ollivier and Zurek proposed a measure called quantum discord \cite{Ollivier,Henderson,Streltsovdiscord}
and stipulated classical correlated states as the states with zero quantum discord.
After that, quantum discord have been widely studied both theoretically and experimentally   \cite{Ali,Girolami,Modi,Daki,Lang,Mazzola,LiB,Maziero1,Werlang,Fanchini,Ferraro,Luo1,Luo2,Luo3}.
Coherence, which marks the departure of quantum theory from classical physics,
is based on the quantum superposition and closely connected to the quantum correlations. Recently, a rigorous framework has been proposed to quantify coherence \cite{Baumgratz},
and the resource theory of coherence has received a great deal of attention \cite{Baumgratz,Winter,Fanheng1,Streltsov,shi}.

A state with zero quantum discord could have non-zero coherence.
Coherence is so general and fundamental that it has many characteristics which are specific for other quantum resources, including quantum discord.
It is worth noting that coherence is a basis-dependent quantifier and can exist in single-partite quantum system,
whereas quantum discord is basis-independent and appears in bipartite quantum system at least.
Although coherence and quantum discord are different,
they are closely connected with each other.
The relationship between quantum discord and coherence is important for us to understand
the common feature of quantum resources \cite{ZWLiu,Yao,MaJJ,TanKC,Guoyu}.
In this paper, we study the relation between quantum discord and coherence
by decreasing the difference between them, i.e., relax the basis constraint $b$ and  remove the local coherence.

Coherence and quantum discord are both a part of total correlation (quantum mutual information),
and coherence is more basic quantum resource.
A nature question is that, the total quantum correlation is characterized by coherence or quantum discord.
There are two views about this question:
i). Coherence indicates the quantum correlation, discord is a special kind of quantum correlation, just like entanglement.
ii). Discord indicates the quantum correlation, coherence contains quantum correlation and a part of classical correlation.
To answer this question, we consider the role of coherence during quantum discord and classical correlation decoherence.

The interaction between a quantum system and its environment
reveals abundant characters of quantum physics,
such as frozen \cite{Bromley} and sudden death \cite{YuTing1,YuTing2} of quantum resources.
We investigate bipartite quantum system evolution under incoherent quantum channels.
It is shown that coherence can both indicate the behavior of quantum discord and classical correlation in different times,
and the role of coherence changes suddenly at transition time $\bar{t}$.
This phenomenon shows that coherence captures both quantum and classical features.

The paper is organized as follows. In Sec. \ref{aaa}, we review the measures of coherence and quantum discord,
and discuss the relationship between coherence and quantum discord measures.
In Sec. \ref{bbb}, we verify that, for Bell-diagonal states, quantum discord is equal to coherence in optimal basis.
In Sec. \ref{ccc}, we study the role of coherence in quantum discord and classical correlation decoherence.
In Sec. \ref{ddd}, we show that coherence frozen and decay indicate the frozen and decay of quantum discord and classical correlation in different times.
We summarize our conclusion and future research in Sec. \ref{eee}.

\

\

\section{measures of quantum correlation}\label{aaa}

\subsection{Measures of coherence}

A reasonable measure to quantify coherence should fulfill \cite{Baumgratz}:
Nonnegativity,  $\mathcal{C}(\rho)\geq0$ with equality if and only if $\rho$ is incoherent;
Monotonicity,  $\mathcal{C}$ do not increase under the action of incoherent operations, $\mathcal{C}(\Lambda[\rho])\leq \mathcal{C}(\rho)]$, for any incoherent operation $\Lambda$;
Strong monotonicity,  $\mathcal{C}$ do not increase on average under selective incoherence operations, $\sum_i q_i \mathcal{C}(\sigma_i)\leq \mathcal{C}(\rho)$, with probabilities $q_i=Tr[K_i\rho K_i^\dag]$, $\sigma_i=K_i\rho K_i^\dag/q_i$, and incoherent Kraus operators $K_i$;
Convexity, $\mathcal{C}$ is a convex function of the state, $\sum_ip_i\mathcal{C}(\rho_i)\geq\mathcal{C}(\sum p_i \rho_i)$.
In accordance with the set of
properties which every proper measure of coherence should
satisfy, a number of coherence measure have been put forward.
We focus on the relative entropy of coherence and $l_1$ norm of coherence.
The relative entropy of coherence
is defined as \cite{Baumgratz}
\begin{equation}\label{}
  \mathcal{C}_{r}(\rho)=\min_{\delta\epsilon\mathcal{I}}S(\rho\|\delta)=S(\rho_{{diag}})-S(\rho),
\end{equation}
where $\rho_{diag}$ comes from $\rho$ by vanishing off-diagonal elements, $\mathcal{I}$ stands for the set of incoherent states, $S(\rho\|\delta)=Tr(\rho\log\rho-\rho\log\delta)$ is the quantum relative entropy and $S(\rho)=-Tr(\rho\log\rho)$ is the von Neumann entropy \cite{Nielsen}.
The $l_1$ norm of coherence
is defined as \cite{Baumgratz}
\begin{equation}\label{}
  \mathcal{C}_{l_1}(\rho)=\min_{\delta\epsilon\mathcal{I}}|\rho-\delta|_{l_1} =\sum_{i\neq j}|\rho_{ij}|,
\end{equation}
where $\rho_{ij}$ are entries of $\rho$.

\subsection{Measures of quantum discord}

The quantum mutual information of system $A$ and $B$ is defined as
\begin{equation}\label{}
  \mathcal{I}(\rho_{AB})=S(\rho_A)+S(\rho_B)-S(\rho_{AB}).
\end{equation}
The one-side classical mutual information is given by the following form
\begin{equation}\label{a}
  \mathcal{J}_{cq}(\rho_{A:B})=S(B)-S(B|\{\Pi_a\}),
\end{equation}
where $S(B|\{\Pi_a\})=\sum_ap_aS(\rho_{B|a})$ is the conditional entropy \cite{Nielsen},
and $\{\Pi_a\}$ is a set of projective measurement with the classical outcome $a$ on subsystem $A$.
The minimized difference between quantum mutual information and one-side classical mutual information

\begin{equation}\label{}
  \D_1(\rho_{AB})\equiv\min_{\{\Pi_a\}}\{\I(\rho_{A:B})-\mathcal{J}_{cq}(\rho_{A:B})\},
\end{equation}
was called quantum discord by Olliver and Zurek \cite{Ollivier}.
The classical correlation was proposed by Henderson and Vedral \cite{Henderson}
\begin{equation}\label{}
  \CC_{cq}(\rho_{AB})\equiv\max_{\Pi_a}\mathcal{J}_{cq}(\rho_{A:B}),
\end{equation}
where the maximum is taken over the complete set of projective measurement $\{\Pi_a\}$.
Then quantum discord is simply defined as $\mathcal{D}_1(\rho_{AB})=\I-\CC(\rho_{AB})$.

One can also define quantum and classical correlations via two-side measurement.
The two-side classical correlation in a composite bipartite system can be expressed as the maximum classical mutual information
\begin{equation}\label{}
  \CC_{cc}(\rho_{AB})\equiv\max_{\Pi_a\otimes \Pi_b}\mathcal{I}_{c}(\rho_{A:B}),
\end{equation}
where $\{\Pi_a\otimes \Pi_b\}$ is a set of local projective measurements with the classical outcome $a$ and $b$ on subsystem $A$ and $B$, $\mathcal{I}_{c}(\rho_{A:B})=\h(\rho_A)+\h(\rho_A)-\h(\rho_{AB})$ is classical mutual information and $\h=\sum_i p_i\log p_i$ is Shannon entropy.
The two-side quantum discord is defined as \cite{Piani1,Piani2}
\begin{equation}\label{}
  \D_2(\rho_{AB})=\I(\rho_{AB})-\CC_{cc}(\rho_{AB}).
\end{equation}
The relative entropy of discord is defined as \cite{Modi}
\begin{equation}
\D_r(\rho_{AB})=\min_{\delta}S(\rho_{AB}\parallel\chi)=\min_{\mathcal{B}(\overrightarrow{k})}\h({\mathcal{B}(\overrightarrow{k})})-S(\rho_{AB}),
\end{equation}
where $\chi$ is in the set of classical-classical states and $\{\ket{\mathcal{B}(\overrightarrow{k})}=\ket{\mathcal{B}(k_1)}\ket{\mathcal{B}(k_2)}$ is a local orthogonal basis.

\subsection{Relation of quantum discord and coherence}

Coherence and discord in a
given state is the distance to the closest incoherent state and classical correlated states.
The relative entropy of quantum discord and relative entropy of coherence are defined as \cite{ModiK,Baumgratz}
\begin{equation}\label{discord}
  \mathcal{D}=\min_{\delta\subset\mathcal{CC}}S(\rho\parallel\delta),
\end{equation}
\begin{equation}\label{coherence}
  \mathcal{C}=\min_{\delta\subset\mathcal{I}}S(\rho\parallel\delta),
\end{equation}
where $\mathcal{CC}$ and $\mathcal{I}$ stand for the sets of classically correlated states (classical-classical states for bipartite system) and incoherent states.
Incoherent states take the form \cite{Baumgratz}
\begin{equation}\label{}
  \delta=\sum_kp_k\ket{b_k}\bra{b_k},
\end{equation}
where $\ket{b_k}=\ket{b_{k,1}}\ket{b_{k,2}}\cdots \ket{b_{k,n}}$.
As for this, the incoherent states can be rewritten as the form of classically correlated states
\begin{equation}\label{incoherentstates}
  \delta=\sum_kp_k\tau_{k,1}^{(b)}\otimes\cdots\otimes\tau_{k,n}^{(b)},
\end{equation}
where $\ket{b_{k,n}}$ which satisfies $\langle b_{k,n}\ket{b_{k\prime,n}}=\delta_{k,k\prime}$ and $\tau_{k,n}^{(b)}=\sum_kp_{k,n}\ket{b_{k,n}}\bra{b_{k,n}}$  are a fixed particular basis $b$ and incoherent state on the subsystem $n$ respectively.
It is necessary to emphasize that although the classically correlated states and incoherent states take the same form, the sets $\mathcal{CC}$ and $\mathcal{I}$ are different collection of the states expressed in Eq. (\ref{incoherentstates}).
They are the collection of the states $\delta$ with no constrain on basis and the states $\delta$ in specific basis,
and the inclusion of sets clearly appears $\mathcal{CC}\supset\mathcal{I}$.

Coherence is a basis-dependent quantity, different basis generate different coherence, i.e., the coherence must be studied in a specific basis.
Whereas, quantum discord is not a basis-depend quantity.
We attempt to give an alternative understanding of how quantum discord is related to
quantum coherence.
To reduce the difference between quantum discord and coherence,
we relax the basis constraint $b$ and minimize over the
set of measurement basis.
Denote the basis $b^{opt}$ is the minimum solution of quantum discord,
we define relative entropy of coherence in basis $b^{opt}$ as $\C_r^{opt}$.
The relation between quantum discord and coherence can be expressed as follows \cite{Yao}
\begin{theorem}\label{bb}
The relative entropy of quantum discord is equal to relative entropy of coherence in an optimal basis.
\end{theorem}

\emph{Proof:}
\begin{eqnarray}
  \D_r(\rho_{AB}) &=& \min_{\mathcal{B}(\overrightarrow{k})}\{\h({\mathcal{B}(\overrightarrow{k})})-S(\rho_{AB})\} \n
   &=& \min_{\mathcal{B}(\overrightarrow{k})}\{S(\rho_{AB_{diag}}^{\mathcal{B}(\overrightarrow{k})})-S(\rho_{AB})\}  \n
   &=& S(\rho_{AB_{diag}}^{opt})-S(\rho_{AB})\n
   &=& \Cr^{opt}(\rho_{AB}),
\end{eqnarray}
where $\rho_{AB_{diag}}^{\mathcal{B}(\overrightarrow{k})}$ and $\rho_{AB_{diag}}^{opt}$ comes from the state $\rho_{AB}$ in basis $\mathcal{B}(\overrightarrow{k})$ and $b^{opt}$ by vanishing off-diagonal elements respectively.

Quantum discord exist at least in two-partite system, but coherence can even exist in one-partite system.
We may remove the local coherence and choose an optimal basis, and then, the relation between
two-side quantum discord and relative entropy of coherence can be expressed as follows
\begin{theorem}
Two-side quantum discord is equal to relative entropy of coherence between subsystem $A$ and $B$ in an optimal basis.
\end{theorem}

\emph{Proof:}
\begin{eqnarray}
  \D_2(\rho_{AB}) &=& \min_{\Pi_a\otimes\Pi_a}\{\I(\rho_{AB})-\I_{cc}(\rho_{AB})\} \n
   &=& \min_{\Pi_a\otimes\Pi_a}\{\Cr(\rho_{AB})-\Cr(\rho_{A})-\Cr(\rho_{B})\} \n
   &=& \Cr^{opt}(\rho_{AB})-\Cr^{opt}(\rho_{A})-\Cr^{opt}(\rho_{B}).
\end{eqnarray}

\section{Coherence and quantum discord for two-qubit system}\label{bbb}



\begin{figure}[ht]
  \centering
  \includegraphics[width=0.49\textwidth]{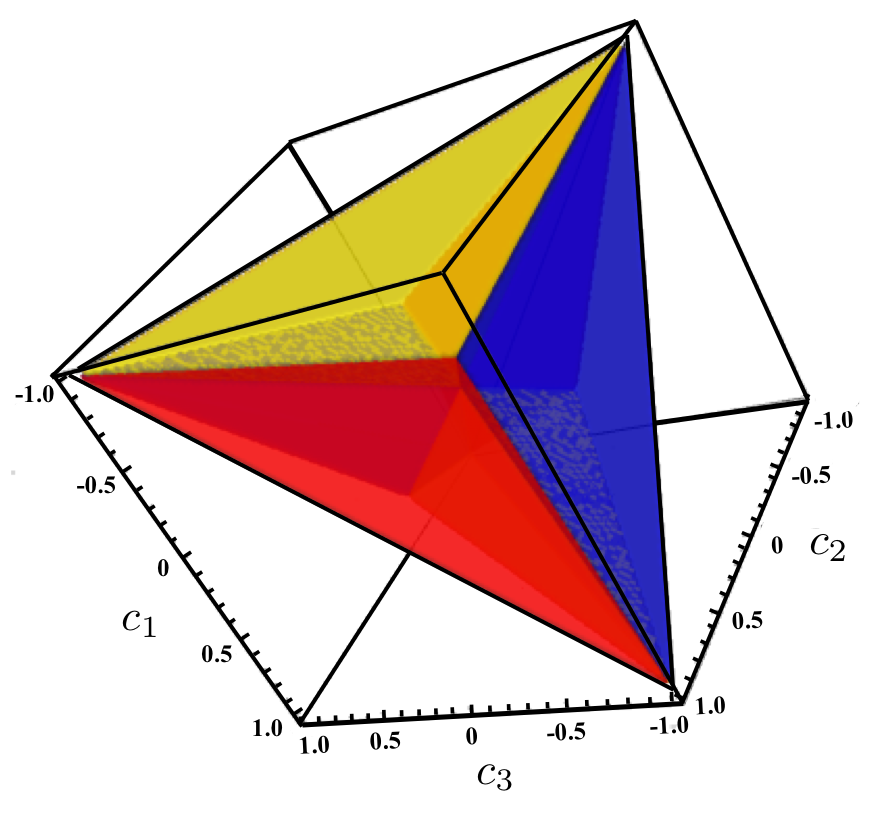}\\
  \caption{According to quantum discord and classical correlation, Bell-diagonal states in the tetrahedron can be divided into three reigns.
  The sudden change of quantum discord and classical correlation appears in the contact area of those three area.
  In yellow reign, $c=c_1$ and $\D(\rho_{AB})=\C(\rho_{AB})^{\sigma_1}$.
  In red reign, $c=c_2$ and $\D(\rho_{AB})=\C(\rho_{AB})^{\sigma_2}$.
  In blue reign, $c=c_3$ and $\D(\rho_{AB})=\C(\rho_{AB})^{\sigma_3}$.
  As for Werner states, $\Cr(\rho_{AB}^{\sigma_1})=\Cr(\rho_{AB}^{\sigma_2})=\Cr(\rho_{AB}^{\sigma_3})=\D(\rho_{AB})$.}\label{bellstatestu}
\end{figure}

The most general two-qubit states $\rho_{AB}$ can be expressed as \cite{Fano}
\begin{equation}\label{}
  \rho_{AB}=\frac{1}{4}(I\otimes I+\vec{a}\vec{\sigma}\otimes I+I\otimes\vec{b}\vec{\sigma}+\sum_{i,j=1}^3c_{ij}\sigma_i\otimes\sigma_j),
\end{equation}
Here $I$ is the identity, $\vec{\sigma}=(\sigma_1,\sigma_2,\sigma_3)$ with $\sigma_1,\sigma_2,\sigma_3$ being the Pauli operator.   $\vec{a},\vec{b}$ are vectors in $\mathbb{R}^3$, and $c_{i,j}$ are real number.
For simplicity,
we will only consider the Bell-diagonal states \cite{Horodecki2}, which can be parameterized as
\begin{equation}\label{Bell-diagonal-states}
  \rho_{AB}=\frac{1}{4}(I\otimes I+\sum_{j=1}^3c_j\sigma_j\otimes\sigma_j)=\sum_{ab}\lambda_{ab}\ket{\beta_{ab}}\bra{\beta_{ab}},
\end{equation}
with the maximally mixed marginals ($\rho_A=\rho_B=\frac{I}{2}$).
The density matrix of Bell-diagonal states with $\sigma_3$ representation takes the form
\begin{equation}\label{}
  \rho_{AB}^{\sigma_z}=\frac{1}{4}\left(
    \begin{array}{cccc}
      1+c_3 & 0 & 0 & c_1-c_2 \\
      0 & 1-c_3 & c_1+c_2 & 0 \\
      0 & c_1+c_2 & 1-c_3 & 0 \\
      c_1-c_2 & 0 & 0 & 1+c_3 \\
    \end{array}
  \right),
\end{equation}\label{Bell}
The eigenstates of $\rho_{AB}^{\sigma_3}$ are the
four Bell states: \qquad\quad\
\begin{equation}\label{}
  \ket{\beta_{ab}}=(\ket{0,b}+(-1)^a\ket{1,1\oplus b})/\sqrt{2},
\end{equation}
and the corresponding eigenvalues are
\begin{equation}\label{}
  \lambda_{ab}=\frac{1}{4}[1+(-1)^ac_1-(-1)^{a+b}c_2+(-1)^bc_3],
\end{equation}
where $a,b\in\{0,1\}$.
The Bell-diagonal states are a three-parameter set, whose geometry
can be depicted as a tetrahedron $\mathcal{T}$ in three-parameter space, see Fig. \ref{bellstatestu}.

Recall that two orthonormal basis
sets $A$ and $B$, for a \textit{d}-dimensional Hilbert space, are said to
be mutually unbiased base, or maximally complementary, if their
overlaps are constant, if $|\bra{A}B\rangle|=d^{-1}$ for all $a$ and
$b$.
Three Pauli qubit observables $\sigma_1$, $\sigma_2$, and $\sigma_3$
are mutually unbiased, in the sense that the distribution
of any one of these observables is uniform for any eigenstate
of the others \cite{Schwinger,Ivanovic,WoottersWK,ChengSM}.
The Bell-diagonal states with $\sigma_x$ and $\sigma_y$ representation take the forms of
\begin{equation}\label{}
  \rho_{AB}^{\sigma_1}=\frac{1}{4}\left(
    \begin{array}{cccc}
      1+c_1 & 0 & 0 & c_3-c_2 \\
      0 & 1-c_1 & c_3+c_2 & 0 \\
      0 & c_3+c_2 & 1-c_1 & 0 \\
      c_3-c_2 & 0 & 0 & 1+c_1 \\
    \end{array}
  \right),
\end{equation}\label{}
and
\begin{equation}\label{}
  \rho_{AB}^{\sigma_2}=\frac{1}{4}\left(
    \begin{array}{cccc}
      1+c_2 & 0 & 0 & c_3-c_1 \\
      0 & 1-c_2 & c_1+c_3 & 0 \\
      0 & c_1+c_3 & 1-c_2 & 0 \\
      c_3-c_1 & 0 & 0 & 1+c_2 \\
    \end{array}
  \right).
\end{equation}\label{Bell}
For Bell-diagonal states, there are no coherence in subsystem. The distribution of coherence is simple and just between two subsystem. The relative entropy of coherence is given by
\begin{eqnarray}
  \mathcal{C}_{r}(\rho_{AB}^{\sigma_i}) =-H(\lambda_{ab})-\sum_{j=1}^2\frac{(1+(-1)^jc_i)}{2}\log_2\frac{(1+(-1)^jc_i)}{4},\n
\end{eqnarray}
where $H(\lambda_{ab})=-\sum_{ab}\lambda_{a,b}\log_2\lambda_{ab}$.
The $l_1$ norm of coherence is
\begin{equation}
  \mathcal{C}_{l_1}=\frac{1}{2}|c_1-c_2|+\frac{1}{2}|c_1+c_2|.
\end{equation}
The mutual information for Bell-diagonal states is given by
\begin{equation}\label{}
  \mathcal{I}=\sum_{a,b}\lambda_{ab}\log_2(4\lambda_{ab}).
\end{equation}
The classical correlation for Bell-diagonal states is given by
\begin{equation}\label{}
  \mathcal{CC}=\sum_{j=1}^2\frac{(1+(-1)^jc)}{2}\log_2(1+(-1)^jc),
\end{equation}
where $c=\max\{|c_1|,|c_2|,|c_3|\}$.
The quantum discord for Bell-diagonal states is given by \cite{Luo2}
 \begin{eqnarray}
  \mathcal{D}(\rho_{AB})=-H(\lambda_{ab})-\sum_{j=1}^2\frac{(1+(-1)^jc)}{2}\log_2\frac{(1+(-1)^jc)}{4}.\n
\end{eqnarray}
We note that one-side quantum discord, two-side quantum discord and relative entropy of quantum discord are identical for Bell-diagonal states.
It is easy to verify that quantum discord is equal to coherence with an optimal basis, and the sudden change of optimal basis comes from the sudden change
of classical correlation. We show that the relation between quantum discord and coherence intuitively in Tab \ref{quantumdiscords}.
\begin{table}[ht]
\caption{As for Bell-diagonal states, quantum discord is equal to coherence with optimal basis.
}
\begin{tabular}{ccccc}
  \hline
  \hline
  \quad region  \qquad        &\qquad $c=c_1$ & \qquad$c=c_2$ & \qquad$c=c_3$\qquad\\
  \hline
  quantum discord & \qquad$\Cr(\rho_{AB}^{\sigma_1})$ &\qquad$\Cr(\rho_{AB}^{\sigma_2})$ &\qquad$\Cr(\rho_{AB}^{\sigma_3})$\qquad\\
  \hline
  \hline
\end{tabular}
\label{quantumdiscords}
\end{table}

\section{Correlations decoherence}\label{ccc}
If no instructions, we choose $\sigma_z$ representation for Bell-diagonal states.
In this section, we study the quantum discord, classical correlation and coherence decoherence under phase flip channel.
Phase flip channel has operation elements \cite{Nielsen}
\begin{eqnarray}
  K_{20} &=& \sqrt{1-q(t)/2}I, \n
  K_{21} &=& \sqrt{q(t)/2}\sigma_3.
\end{eqnarray}
where $q=e^{-2\gamma t}$ with damping rate $\gamma$ is noisy strength. We put phase flip channel on system $A$ and system $B$ respectively, the time evolution of Bell-diagonal states can be expressed as
\begin{eqnarray}
  c_1(t) &=& c_1(0)e^{-2\gamma t}, \n
  c_2(t) &=& c_2(0)e^{-2\gamma t}, \n
  c_3(t) &\equiv& c_3(0).
\end{eqnarray}
For the states $c_3=\max\{|c_1|,|c_2|,|c_3|\}$, $\mathcal{C}_{r}(\rho_{AB})=\mathcal{D}(\rho_{AB})$,
For the states $c_1=\max\{|c_1|,|c_2|,|c_3|\}$, $\mathcal{C}_{r}(\rho_{AB})>\mathcal{D}(\rho_{AB})$, $\CC(\rho_{AB})=H(C_{l_1})$
for the times $t<\bar{t}_1=-\frac{\ln c_3(0)-\ln c_1(0)}{2\gamma} $,
and $\mathcal{C}_{r}(\rho_{AB})=\mathcal{D}(\rho_{AB})$
for the times $t>\bar{t}_1=-\frac{\ln c_3(0)-\ln c_1(0)}{2\gamma} $.
For the states $c_2=\max\{|c_1|,|c_2|,|c_3|\}$, $\mathcal{C}_{r}(\rho_{AB})>\mathcal{D}(\rho_{AB})$, $\CC(\rho_{AB})=H(C_{l_1})$
for the times $t<\bar{t}_2=-\frac{\ln c_3(0)-\ln c_2(0)}{2\gamma} $,
and $\mathcal{C}_{r}(\rho_{AB})=\mathcal{D}(\rho_{AB})$
for the times $t>\bar{t}_2=-\frac{\ln c_3(0)-\ln c_2(0)}{2\gamma}$.
As $c_3$ close to zero, $\bar{t}$ is increasing exponentially.
While $c_3\rightarrow0$, $\bar{t}\rightarrow\infty$, coherence describe the behavior of classical correlation all the time.
Coherence describes the behavior of classical correlation for times $t<\bar{t}$, but coherence describe the behavior of quantum discord for times $t>\bar{t}$.
The sudden change also comes from the sudden change of quantum discord and classical correlation.
The trajectory of time evolution from red or yellow reign to blue reign at $\bar{t}$,
and the time $\bar{t}$ is the time where the role of coherence sudden change.
In Tab. \ref{discordcoherence}, We show the relation of quantum discord, classical correlation and coherence in different times.
\begin{table}[ht]
\caption{Coherence indicates quantum discord and classical correlation decoherence in different times.
}
\begin{tabular}{cccc}
  \hline
  \hline
  \quad times  \qquad        &\qquad $t<\bar{t}$ & \qquad$t>\bar{t}$ \\
  \hline
  classical correlation       & \qquad$H(\C_{l_1})$& \qquad$\I-\Cr$\\
  \hline
  quantum discord   & \qquad$\I-H(\C_{l_1})$ &\qquad$\C_r$ \\
  \hline
  \hline
\end{tabular}
\label{discordcoherence}
\end{table}

\begin{figure}[ht]
  \centering
  \includegraphics[width=0.49\textwidth]{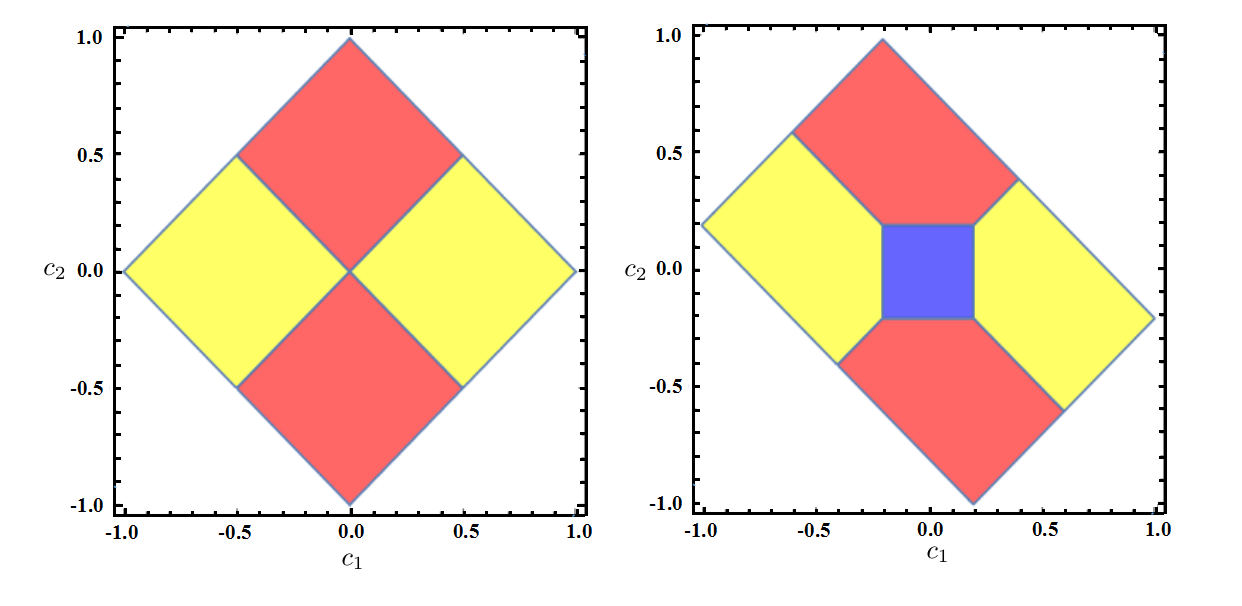}\\
  \caption{The left is the trajectory of states with $c_3=0$, and the right is for the states with $c_3=0.2$. In blue reign, $c=c_3$ and $\D(\rho_{AB})=\C(\rho_{AB})^{\sigma_3}$, and out of blue reign $\CC=\mathcal{C}_{l_1}$. The trajectory of time evolution of Bell-diagonal states under phase flip channel is close to $c_3$ axis along straight line.  The trajectory of time evolution for Bell-diagonal states will get to blue reign over time $\bar{t}$, and the role of coherence sudden change at time $\bar{t}$.}\label{}
\end{figure}

\

The fact that coherence describe the behavior of quantum discord and classical correlation in different times
also exist in $\sigma_1$ and $\sigma_2$ representation,
just replace phase flip channel by bit flip and bit-phase flip channel.

\section{The behavior of correlations indicated by coherence}\label{ddd}
Sudden transition between classical correlation and
quantum discord loss in a composite system
has been studied in \cite{Mazzola}.
We are going to study the role of coherence in sudden transition between classical correlation and quantum discord loss.

\subsection{Coherence indicates the frozen of quantum discord and classical correlation}

\begin{figure}[ht]
  \centering
  \includegraphics[width=0.49\textwidth]{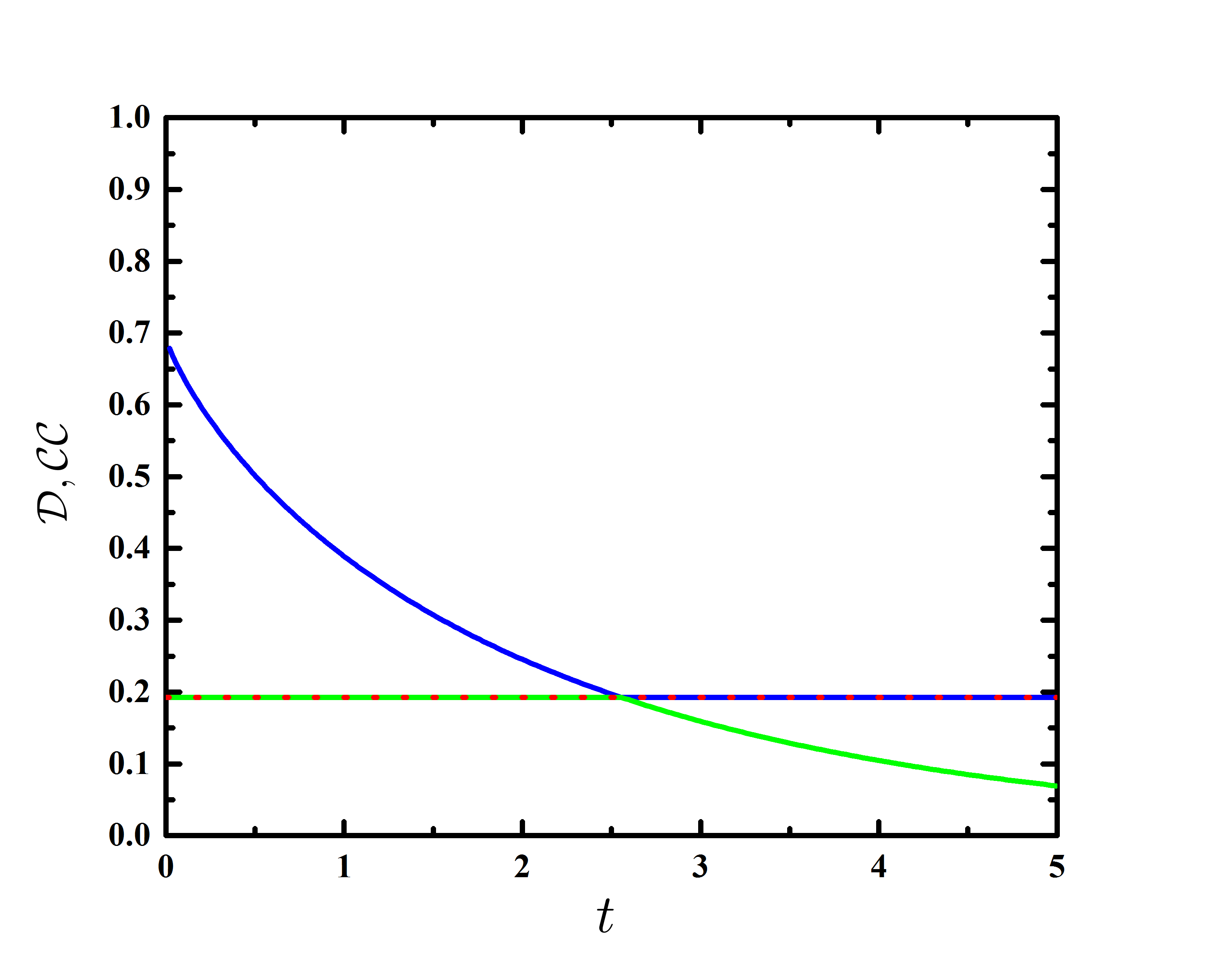}\\
  \caption{For the initial states $c_1(0)=0.6$, $c_2(0)=-0.6$, $c_3(0)=1$ and $\gamma=0.1$, coherence (red dots) shows the frozen phenomenon of quantum discord (green line) and classical correlation (blue line) under bit flip channel.}\label{frozen}
\end{figure}

The \textit{bit flip} channel flip the state of qubit from $\ket{0}$ to $\ket{1}$ (and vice versa) with noisy strength $q$.
It has operation elements \cite{Nielsen}
\begin{eqnarray}
  K_{10} &=& \sqrt{1-q(t)/2}I \n
  K_{11} &=& \sqrt{q(t)/2}\sigma_1.
\end{eqnarray}
We put bit flip channel on system $A$ and system $B$ respectively, the time evolution of Bell-diagonal states can be expressed as
\begin{eqnarray}
  c_1(t) &\equiv& c_1(0), \n
  c_2(t) &=& c_2(0)e^{-2\gamma t}, \n
  c_3(t) &=& c_3(0)e^{-2\gamma t}.
\end{eqnarray}
For the initial states $c_2(0)=-c_1(0)c_3(0)$, the coherence is frozen \cite{Bromley}.
For the initial states $c_2(0)=-c_1(0)c_3(0)$, $c\neq c_1$
the quantum discord is frozen but the classical correlation is decreasing before the transition time $\overline{t}$,
and the quantum discord is decreasing but the classical correlation is frozen over the transition time $\overline{t}$.
For the initial states $c_2(0)=-c_1(0)c_3(0)$, $c=c_1$, coherence shows the frozen of classical correlation for all time.
Coherence shows the frozen phenomenon of quantum discord and classical correlation.
In Fig. \ref{frozen}, we plot the time evolution of the quantum discord, the classical correlations,
and the coherence for $c_1(0)=0.6$, $c_2(0)=-0.6$, $c_3(0)=1$ and $\gamma=0.1$.
The plot clearly shows the frozen of coherence indicates the frozen of quantum discord and classical correlation in different times.
The action of bit-phase flip channel on initial states $c_2(0)=-c_1(0)c_3(0)$ also indicated
that coherence shows the frozen phenomenon of quantum discord and classical correlation.

\subsection{Coherence indicates the decay of quantum discord and classical correlation}
\begin{figure}[ht]
  \centering
  \includegraphics[width=0.49\textwidth]{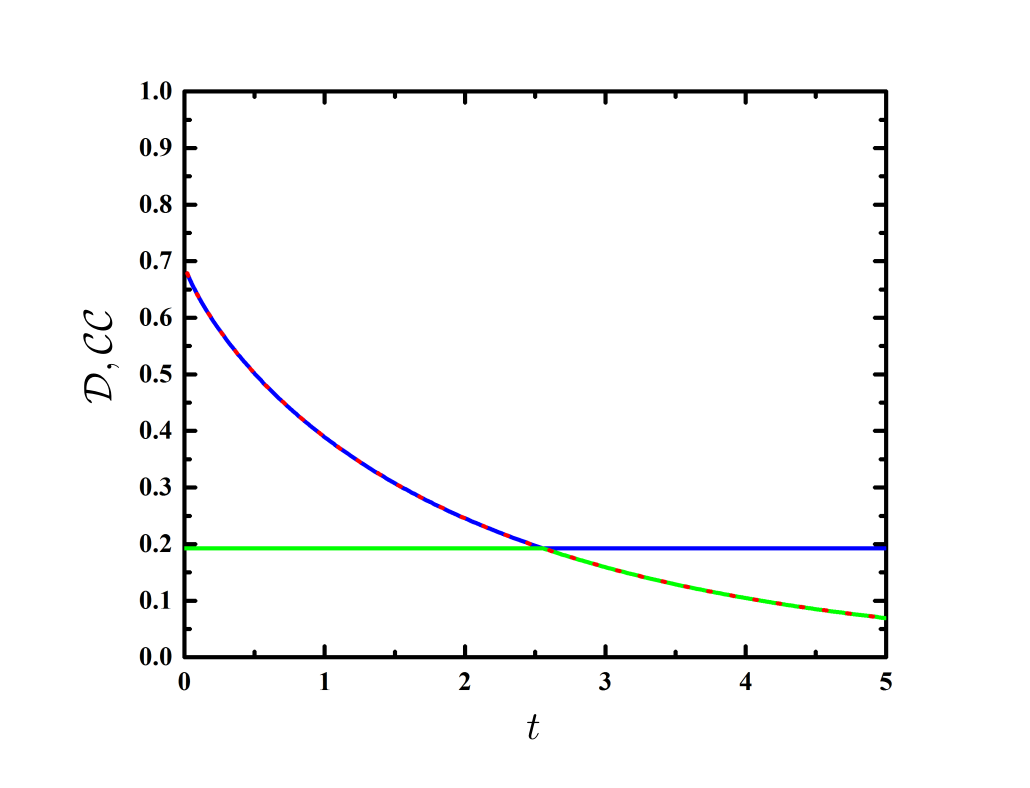}\\
  \caption{For the initial states $c_1(0)=1$, $c_2(0)=-0.6$, $c_3(0)=0.6$ and $\gamma=0.1$, coherence (red dots) shows the decreasing phenomenon of quantum discord (green line) and classical correlation (blue line).}\label{decay}
\end{figure}

The \textit{phase flip} channel has operation elements \cite{Nielsen}
\begin{eqnarray}
  K_{20} &=& \sqrt{1-q(t)/2}I, \n
  K_{21} &=& \sqrt{q(t)/2}\sigma_3.
\end{eqnarray}
We put phase flip channel on system $A$ and system $B$ respectively, the time evolution of Bell-diagonal states can be expressed as
\begin{eqnarray}
  c_1(t) &=& c_1(0)e^{-2\gamma t}, \n
  c_2(t) &=& c_2(0)e^{-2\gamma t}, \n
  c_3(t) &\equiv& c_3(0).
\end{eqnarray}
For the initial states $c_2(0)=-c_1(0)c_3(0)$, $c\neq c_3$,
coherence shows the decay of classical correlation for times $t<\bar{t}$,
and shows the quantum discord decay for times $t>\bar{t}$.
For the initial states $c_2(0)=-c_1(0)c_3(0)$, $c= c_3$, coherence shows the decay of quantum discord for all times.
In Fig. \ref{decay}, we plot the time evolution of the quantum discord, the classical correlations,
and the coherence for $c_1(0)=1$, $c_2(0)=-0.6$, $c_3(0)=0.6$ and $\gamma=0.1$.
The plot clearly shows the decay of coherence indicates the decay of quantum discord and classical correlation in different times.
The action of phase damping channel on initial states  also indicated that coherence shows
the decay of quantum discord and classical correlation.

\section{Conclusion}\label{eee}
In this paper, we have studied the role of coherence in quantum discord and classical correlation decoherence.
In order to do this issue, we first study the relationship between coherence and quantum discord by reducing the difference between them.
Relax the basis constraint $b$, the relative entropy of quantum discord is equal to the relative entropy of coherence in optimal basis.
Remove the local coherence in subsystems, the two-side quantum discord is equal to the coherence between two subsystems in optimal basis.

As for Bell-diagonal states, quantum discord is equal to coherence in a set of MUBs.
And then, we study the dynamics of quantum discord, classical correlation and coherence for Bell-diagonal states.
It is shown that coherence describes the behavior of classical correlation and quantum discord for times $t<\bar{t}$ and $t>\bar{t}$ respectively.
The role of coherence during classical and quantum decoherence changes at transition time $\bar{t}$ suddenly, this phenomenon comes from the sudden change of the optimal basis.
Moreover, the coherence frozen and decay shows the frozen and decay of quantum discord and classical correlation in different times, respectively.

We believe that our work is important to understand the common features of quantum resources and the role of quantum discord, classical correlation and coherence
in quantum resources theory. In future works, we will study the essential relations between quantum resources and try to provide a unified framework of them.

\acknowledgments
We thank Feng-Lin Wu, Hai-Long Shi and Wei Xia for their valuable discussions. This work was
supported by the NSFC (Grants No. 11375141, No. 11425522, No. 91536108, No. 11647057 and No. 11705146),
the Special Research Funds of Shaanxi Province Department of Education (No. 203010005), Northwest University Scientific Research Funds (No. 338020004) and the Double First-Class University Construction Project of Northwest University.

\end{document}